\documentclass[manuscript,screen]{acmart}

\usepackage{listings}
\definecolor{codegreen}{rgb}{0,0.6,0}
\definecolor{codegray}{rgb}{0.5,0.5,0.5}
\definecolor{codepurple}{rgb}{0.58,0,0.82}
\definecolor{backcolour}{rgb}{0.95,0.95,0.92}
\lstdefinestyle{myStyle}{
    belowcaptionskip=1\baselineskip,
    breaklines=true,
    frame=none,
    numbers=left,
    basicstyle=\footnotesize\ttfamily,
    keywordstyle=\bfseries\color{green!40!black},
    commentstyle=\itshape\color{purple!40!black},
    identifierstyle=\color{blue},
}
\usepackage{multirow}
\usepackage{tabularx}
\usepackage{caption}
\usepackage{subcaption}

\setcopyright{acmcopyright}
\copyrightyear{2022}
\acmYear{2022}
\acmDOI{XXXXXXX.XXXXXXX}

\acmConference[IWOCL and SYCLcon '22]{...}{May 10--12, 2022}{Virtual}
\acmPrice{15.00}
\acmISBN{978-1-4503-XXXX-X/18/06}

\begin{document}

\title[A Proof-of-Concept SYCL FFT]{Benchmarking a Proof-of-Concept Performance Portable SYCL-based Fast Fourier Transformation Library}

\author{Vincent R. Pascuzzi}
\email{pascuzzi@bnl.gov}
\orcid{1234-5678-9012}
\affiliation{%
  \institution{Brookhaven National Laboratory}
  \streetaddress{P.O. Box 1212}
  \city{Dublin}
  \state{Ohio}
  \country{USA}
  \postcode{43017-6221}
}
\author{Mehdi Goli}
\email{mehdi.goli@codeplay.com}
\orcid{0000-0002-3520-9598}
\affiliation{%
  \institution{Codeplay Software Ltd.}
  \streetaddress{P.O. Box 1212}
  \city{Dublin}
  \state{Ohio}
  \country{UK}
  \postcode{43017-6221}
}

\renewcommand{\shortauthors}{Pascuzzi and Goli}

\begin{abstract}
In this paper, we present an early version of a SYCL-based FFT library, capable of running on all major vendor 
hardware, including CPUs and GPUs from AMD, ARM, Intel and NVIDIA.
Although preliminary, the aim of this work is to seed further developments for a rich set of features for calculating
FFTs.
It has the advantage over existing portable FFT libraries in that it is single-source, and therefore removes the
complexities that arise due to abundant use of pre-process macros and auto-generated kernels to target different architectures.
We exercise two SYCL-enabled compilers, Codeplay ComputeCpp and Intel's open-source
LLVM project, to evaluate performance portability of our SYCL-based FFT on various heterogeneous architectures. 
The current limitations of our library is it supports single-dimension FFTs up to
$2^{11}$ in length and base-2 input sequences.
We compare our results with highly optimized vendor specific FFT libraries and provide a detailed analysis to demonstrate a fair level of performance, as well as potential sources of performance bottlenecks.
\end{abstract}



\maketitle

\section{Introduction}
\label{sec:intro}
The Fast Fourier Transform (FFT) is a widely used algorithm for efficiently
computing the discrete Fourier transforms (DFT) of complex- or real-valued data sequences.
The transformed data can be decomposed into the multiple pure
frequencies that make it up, a technique useful in a wide range of applications---from fault analysis, quality control, and condition monitoring of machines or systems to AI machine learning and deep learning~\cite{10.5555/1941838}.
As such, there are numerous FFT implementations provided by nearly all vendors;
however, these implementations are tied to a single architecture or platform, requiring management of multiple codebases for a
single software framework or application.
Several frameworks exist---such as FFTW~\cite{1386650, Padua2011} and VkFFT~\cite{vkfft}---which embed various FFT APIs through the
use of pre-process macros and vendor-supplied libraries to auto-generate architecture-specific kernels to
target different devices.
While such codes are undoubtedly powerful, they suffer in readability (\textit{e.g.}, extensive use of macros) and maintainability (\textit{e.g.}, multi-language code generation).
Moreover, supporting constantly evolving APIs and their languages, while also providing backward
compatibility, complicates macro- and code-generation-based solutions further.
As an alternative to the above approach, open-standard parallel programming APIs, such as SYCL, can be
used to implement FFT and other algorithms directly to support multiple hardware platforms.
Although performance of vendor-specific APIs is likely to be better in practice, a SYCL-based FFT will provide 
portability and also address the readability and maintainability pitfalls.
Whereas our previous work on linear algebra and random number generation
routines~\cite{9652858, 2021arXiv210901329P} utilized SYCL's
interoperability functionality, here we
instead implement FFT algorithms directly using the SYCL programming model---removing
entirely the reliance on third-party libraries.

\section{Related Work}

There are several implementations of FFTs supporting various architectures.
FFTW is a C based implementation of discrete Fourier transform that has been adopted mainly for CPU use.
However, the code generated by an auto generator tool can potentially make the support of new devices and architectures difficult\footnote{From the Github site: ``YOU WILL BE UNABLE TO COMPILE CODE FROM THIS REPOSITORY unless you have special tools and know what you are doing.''}.
VkFFT provides an open source FFT library for accelerators by providing backend implementation of CUDA, OpenCL, HIP, and Vulkan. However, both the library code itself (a single header totaling nearly
40k lines) and generated codes suffer from code duplication due to the lack of high-level programming
feature support.
In practice, this tends to make code more error prone, less maintainable and more difficult to add or
optimise for new devices.

There are several vendors specified FFT which are been tuned specific architectures but are vendor-locked and not portable to other others.
cuFFT\cite{cuFFT} is a closed-source FFT API that runs only on NVIDIA devices.
The package also provides cuFFTW, a porting tool to enable users of FFTW to leverage NVIDIA GPU
compute capabilities.
oneMKL\cite{oneMKL} also provide a closed-source FFT library that runs on x86\_64
architectures, and AMD provides rocFFT\cite{rocFFT} for computing FFTs on ROCm architectures.

In this paper, we introduce SYCL-FFT, an open- and single-source FFT implementation to target a wide
range of heterogeneous devices.
The development employs modern C++ features, such as template meta-programming supported by SYCL, to
provide a parametric representation of FFT kernels that can abstract out the kernel implementation
from the kernel modification.
Using this approach, it is possible to tune the same kernel for various architectures with
little-to-no modification to the actual kernel implementation.
\section{Fast Fourier Transform}
\label{sec:fft}
Fast Fourier transforms are computed by discretizing the input function, represented in a time or space domain, and
mapping it to an output, represented in its frequency domain.
In closed form, the 1D DFT is written as,
\begin{align}
\label{eqn:dft}
    X_k &= \sum_{n=0}^{N-1} x_n e^{-i2\pi kn/N} = \sum_{n=0}^{N-1} x_n \omega_N^{kn}
\end{align}
where $k = [0, N-1]$, $\{x_i\}$ is the real- or complex-valued input sequence of length $N$, and
$\omega_N = e^{-i2\pi/N}$ is the $N$th de Moivre number (root of unity).
The inverse DFT (iDFT) is obtained simply by changing the sign in the exponent,
\begin{align}
\label{eqn:idft}
    x_k = \frac{1}{N} \sum_{n=0}^{N-1} x_n \omega_N^{-kn},
\end{align}
with $1/N$ being the normalization constant.
From Eqns.~\eqref{eqn:dft}, it can be seen that there are $N$ outputs, $X_k$, each of which requires a sum of $N$ terms.
Thus, a direct (na\"{i}ve) evaluation of the DFT has computational complexity $\mathcal{O}(N^2)$.
\begin{figure}[!ht]
    \centering
    \includegraphics[width=0.7\textwidth]{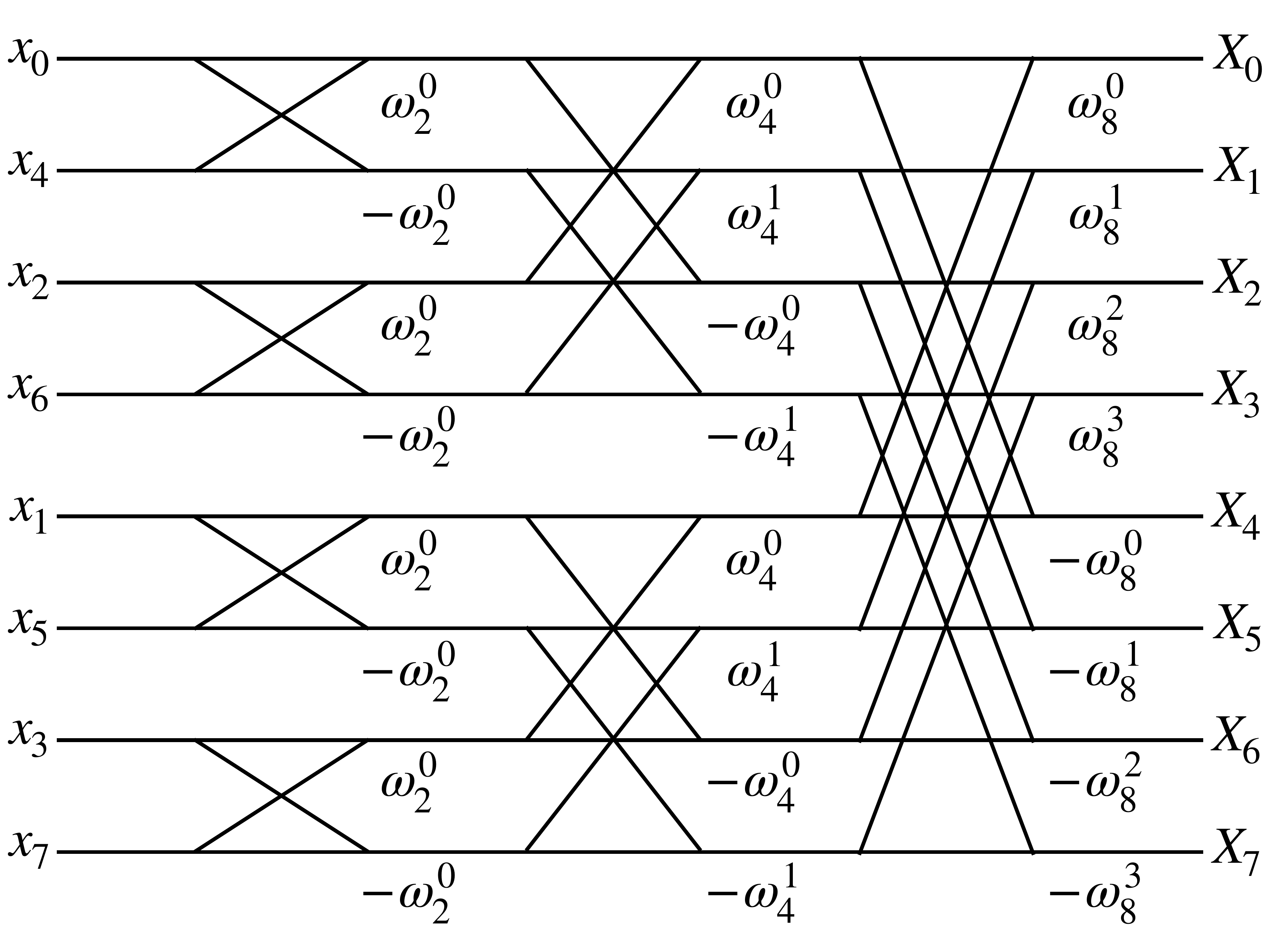}
    \caption{Illustration of a radix-2 DIT on an DFT with input size $N=8$.
                Intersecting vertical lines---displaying butterfly-like patterns---among the input, $\{x_i\}$, correspond to combinations of additions and subtractions as per the twiddle factors,
                $\omega_N^k$; see main text.}
    \label{fig:radix2-8}
\end{figure}
\subsection{Cooley-Tukey algorithm}
There are many FFT algorithms that reduce the complexity of DFT calculations~\cite{10.2307/2310304, 1163036, 1448407, 1162132, 7532670}, the most commonly used being Cooley-Tukey~\cite{cooley1965algorithm}.
Exploiting FFT periodicity, the Cooley-Tukey algorithm uses the divide-and-conquer technique to recursively---or
using a breadth-first traversal of the computational tree---reduce a DFT of composite length
$N$ into smaller DFTs.
The simplest case of the Cooley-Tukey algorithm is the radix-2 decimation in time (DIT, bit order reversal),
illustrated in Figure~\ref{fig:radix2-8} for a length $N=8$ DFT.
In the radix-2 DIT, the input sequence of length $N$ is rearranged into two subsets, one containing all even-indexed sequence elements and the other containing all odd-indexed elements.
Each subset is then separately summed over;
from Eqn.~\eqref{eqn:dft},
\begin{align}
    \label{eqn:radix2}
    X_k &= \sum_{n=0}^{N/2 - 1} x_{2n} \omega_{N/2}^{(2n) k/N}
            + \sum_{n=0}^{N/2 - 1} x_{2n+1} \omega_{N/2}^{(2n+1) k/N} \\
        &= \sum_{n=0}^{N/2 - 1} \left( x_{2n} \omega_{N/2}^{kn}
        + \omega_{N}^k x_{2n+1} \omega_{N/2}^{kn} \right) \\
        &\equiv E_k + \omega_{N}^{k} O_k,
\end{align}
where the first factor includes even ($E$) indices and the second odd ($O$), and the summation is over
$k=[0, N/2-1]$.
Given the periodicity of the complex exponential, the remaining $N/2$ elements can be written similarly,
\begin{align}
    \label{eqn:radix2}
    X_{k+N/2} = \sum_{n=0}^{N/2 - 1} \left( x_{2n} \omega_{N/2}^{kn}
        - \omega_{N}^k x_{2n+1} \omega_{N/2}^{kn} \right),
\end{align}
where the factor of $-1$ in the second term arises from an $e^{-i\pi}$ when simplifying.
Provided the input length $N$ is a power of $2$, the original DFT can be split $\log n$ times.
Since there will still be a sum over $N$ terms for each splitting, the complexity becomes $\mathcal{O}(N \log n)$.

Higher-order radices can reduce the number of arithmetic operations needed to compute larger DFTs (\textit{e.g.}, radix-4, radix-8 and radix-16), as can combinations of different radices
(split-radix decompositions).
A split-radix algorithm reduces a single length $N$ DFT into three smaller summations at each step.
As before, the summations are split into even and odd indices, giving,
\begin{align}
    \label{eqn:split-radix2}
        X_{k} &= \sum_{n_2=0}^{N/2 - 1} x_{2n_2} \omega_{N/2}^{n_2 k}
            + \sum_{n_4=0}^{N/4 - 1}
            \left( \omega_{N}^{k} x_{4n_4+1} \omega_{N/4}^{n_4 k}
             + \omega_{N}^{3k} x_{4n_4 + 3} \omega_{N/4}^{n_4 k} \right) \\
             &\equiv E_{k} + \omega_{N}^{k} O_{k} + \omega_{N}^{3k} O'_{k},
\end{align}
with indices $n_m=N/m - 1$.
Here, the first summation, $E_{k}$, is the even portion of a radix-2 DIT and the second,
$\omega_{N}^{k} O_{k} + \omega_{N}^{3k} O'_{k}$, contains the two odd portions of a radix-4.
The efficiency of this algorithm is due again to the periodicity of $k$;
if we add $N/4$ ($N/2$) to $k$, the radix-4 (radix-2) portions are left unchanged.
Using this fact, we see,
\begin{align}
    \label{eqn:twiddle}
    \omega_{N}^{k+N/4} = -i\omega_{N}^{k}, \\
    \omega_{N}^{3(k+N/4)} = i\omega_{N}^{3k};
\end{align}
\textit{i.e.}, only $\omega_{N}^{k}$ and $\omega_{N}^{3k}$ (so-called \textit{twiddle factors}), need to be updated.
As a result, all output values, $X_k$, in the frequency domain can be calculated via,
\begin{align}
    \label{eqn:split-radix-final}
    X_k &= E_k + \omega_N^k O_k + \omega_N^{3k} O_k' \\
    X_{k+N/2} &= E_k - (\omega_N^k O_k + \omega_N^{3k} O_k') \\
    X_{k+N/4} &= E_{k+N/4} - i(\omega_N^k O_k - \omega_N^{3k} O_k') \\
    X_{k+3N/4} &= E_{k+N/4} + i(\omega_N^k O_k - \omega_N^{3k} O_k')
\end{align}
for $k = [0, N/4]$.
The combinations of additions and subtractions are known as butterflies, depicted by the intersecting
vertical lines in Fig.~\ref{fig:radix2-8}.
\section{Implementation}
\label{sec:impl}
Our SYCL FFT library implements the Cooley-Tukey radix-2 algorithm described above, as well as radix-4, and
radix-8 algorithms.
The class interface is shown in Listing~\ref{lst-fft-interface}.
\begin{center}
\begin{lstlisting}[frame=tb, xleftmargin=.1\textwidth, xrightmargin=0.1\textwidth, language=C++,
caption=SYCL FFT function object interface., label=lst-fft-interface,captionpos=b]
template <typename T, size_t WG_SIZE, int WG_FACTOR, int SYCL_LANGUAGE_VERSION>
class fft_1d {
 public:
  using float2 = sycl::float2;
if constexpr (SYCL_LANGUAGE_VERSION < 202000) {
  using size_accessor = sycl::accessor<size_t, 1, sycl::access::mode::read,
                                       sycl::access::target::global_buffer>;
  using read_accessor = sycl::accessor<T, 1, sycl::access::mode::read,
                                       sycl::access::target::global_buffer>;
  using write_accessor = sycl::accessor<T, 1, sycl::access::mode::write,
                                        sycl::access::target::global_buffer>;
 } else {
  using stage_accessor = sycl::accessor<size_t, 1, sycl::access::mode::read,
                                        sycl::access::target::device>;
  using read_accessor = sycl::accessor<T, 1, sycl::access::mode::read,
                                       sycl::access::target::device>;
  using write_accessor = sycl::accessor<T, 1, sycl::access::mode::write,
                                        sycl::access::target::device>;
}
  using local_rw_accessor = sycl::accessor<T, 1, sycl::access::mode::read_write,
                                           sycl::access::target::local>;
if constexpr (SYCL_LANGUAGE_VERSION >= 202000){
  [[sycl::reqd_work_group_size(WG_SIZE)]]
}
  fft1d(size_accessor stage_sizes, read_accessor inputs,
        write_accessor outputs, local_rw_accessor local_shared,
        size_t work_group_size, int direction);
  void operator()(sycl::nd_item<1> item) const {...}
  inline void radix_2(sycl::nd_item<1> item, size_t stage_mod,
                      float2* temp) const {...}
  inline void radix_4(sycl::nd_item<1> item, size_t stage_mod,
                      float2* temp) const {...}
  inline void radix_8(sycl::nd_item<1> item, size_t stage_mod,
                      float2* temp) const {...}
};
\end{lstlisting}
\end{center}
The SYCL-FFT functor is a templated class requiring three template arguments to define the input data type,
the required work-group size and a constant, \texttt{WG\_FACTOR} that depends on the input sequence length.
Since variable array sizes are not permitted in SYCL kernels, \texttt{WG\_FACTOR} is automatically determined
a priori on the host and the corresponding kernel is called based on this value.
The \texttt{fft1d} class takes five arguments to determine FFT execution;
\texttt{stage\_sizes} is an array of numbers calculated on the host, used to derive the internal steps that
need to be taken (\textit{e.g.}, the sequence of radix function calls;
\texttt{inputs} and \texttt{outputs} are memory allocations on the device for storing the sequence to transform and the transformed output, respectively; \texttt{local\_shared} is
the cross-work-group shared memory; and \texttt{direction} specifies whether to perform an FFT or iFFT
(\texttt{SYCLFFT\_FORWARD} or \texttt{SYCLFFT\_INVERSE}).

It must be noted that our SYCL FFT library is largely a proof-of-concept work-in-progress, and therefore fairly
limited in capability and features.
In its current state, the library can compute 1D single-precision complex-to-complex (C2C) DFTs up to $2^{11}$ in length\footnote{This value is ultimately determined by the number of compute
(execution) units available on a given device.
For example, SYCL-FFT can perform on an input length up to $2^{12}$ on an ARM Neoverse CPU.}.
All transforms are performed out-of-place.
Nevertheless, the aim of this early work is to provide the foundations for an open-source performance
portable FFT library.
%
\section{Experimental Setup}
\label{sec:experiments}
To enable execution across a diverse set of platforms, we have employed two compilers.
To target PTX64, HIP and x86\_64 architectures, Intel open-source LLVM compiler project (referred to
simply as ``Intel LLVM'' in what follows), based on the \texttt{sycl-nightly/20220223} branch based on
LLVM major version 15, was used.
Intel LLVM with support for PTX64 and HIP was built using GCC 10.2.0 and 8.2.0, respectively, with CUDA 11.5.0 and ROCm 4.2.0.
The Portable Compute Language (POCL) used as an OpenCL driver along with Codeplay's ComputeCpp~\cite{} 
to run SYCL-FFT on ARM CPU devices.
The POCL host (where POCL runs) compiler was built using GCC 10.2.0 and its target (where OpenCL runs)
compiler using the Intel open-source LLVM compiler project.
Due to LLVM compatibilty, the POCL target compiler was built with the \texttt{sycl-nightly/20220210}
branch of Intel LLVM, as later nightlies bumped the LLVM major version to 15 which is not yet supported
by POCL.
Details regarding the systems used in these studies are shown in Table~\ref{tab:hw}.
\begin{table*}[t]
    {\centering
    \begin{tabularx}{0.96\columnwidth}{c|c|c|c|c}
 Device&Maximum &\multirow{2}{*}{Backend}&\multirow{2}{*}{Compiler(s)}&\multirow{2}{*}{FFT Library} \\
 (Architecture)&Work-Group Size&&&\\
 \hline
 \hline
  ARM Neoverse-N1 &\multirow{2}{*}{4096} &POCL 1.9&\multirow{2}{*}{ComputeCpp 2.8.0} &\multirow{2}{*}{---} \\
  (ARMv8-A)&&pre-gde9b966b&& \\
 \hline
  Intel Xeon E3-1585 v5&\multirow{2}{*}{8192} &OpenCL 3.0 &\multirow{2}{*}{ComputeCpp 2.8.0}&\multirow{2}{*}{---} \\
  (x86\_64)&&2021.12.9.0.24\_005321&&\\
 \hline
  Intel Iris P580&\multirow{2}{*}{256} &OpenCL 3.0&\multirow{2}{*}{ComputeCpp 2.8.0}&\multirow{2}{*}{---}\\
  (Gen9)&&2021.12.9.0.24\_005321&&\\
 \hline
  AMD MI-100&\multirow{2}{*}{256} &\multirow{2}{*}{HIP 4.2.0}&\texttt{sycl-nightly/20220223}&\multirow{2}{*}{rocfft 4.2.0}\\
  (CDNA)&&&\texttt{hipcc} 4.2.21155 &\\
 \hline
  NVIDIA A100&\multirow{2}{*}{1024} &\multirow{2}{*}{PTX64} &\texttt{sycl-nightly/20220223} &\multirow{2}{*}{cufft 11.5.0} \\
  (Ampere)&&&\texttt{nvcc} 11.5.0 \\
 \hline
    \end{tabularx}
    \caption{Device hardware and software versions for each platform considered in these studies.
            The \texttt{sycl-nightly/x} compilers refer to the specific branch of the Intel LLVM compiler
            project.
            All systems run openSUSE 15.3, kernel version 5.3.18.}
    \label{tab:hw}
    }
\end{table*}
\section{Results}
\label{sec:results}
Without loss of generality, we evaluate our FFT library using the simple linear function $f(x) = x$.
Input sequences in the range $2^3$--$2^{11}$ are produced on the host device and transferred to the device
were the FFT is computed.
All compute nodes were dedicated and not shared among additional users during experimentation.
\subsection{Computational Performance}
The following timing measurements do not include host-side data preparation or data transfers between host
and device.
Each data point corresponds to the mean of 1000 iterations for a given sequence length;
the first launch is considered as a ``warm-up'' and therefore discarded\footnote{The warm-up execution typically is a one-off, an order of magnitude or more larger than
subsequent calculations.}.

\begin{figure}[!ht]
    \begin{subfigure}[b]{0.8\textwidth}
        \centering
        \includegraphics[width=\textwidth]{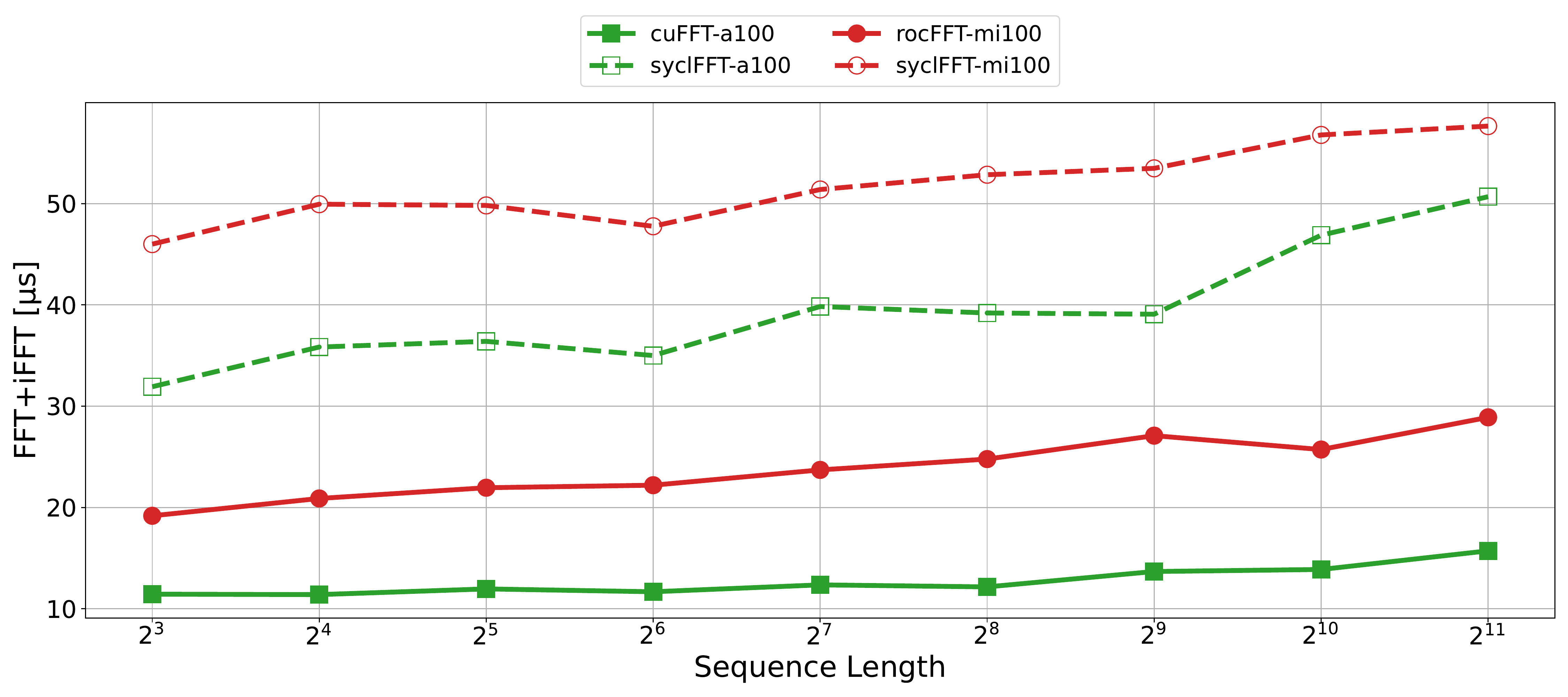}
        \caption{Optimal total (kernel dispatch + execution) run-times.}
        \label{fig:clock-sycl-cuhip-mean}
    \end{subfigure}
    \begin{subfigure}[b]{0.8\textwidth}
        \centering
        \includegraphics[width=\textwidth]{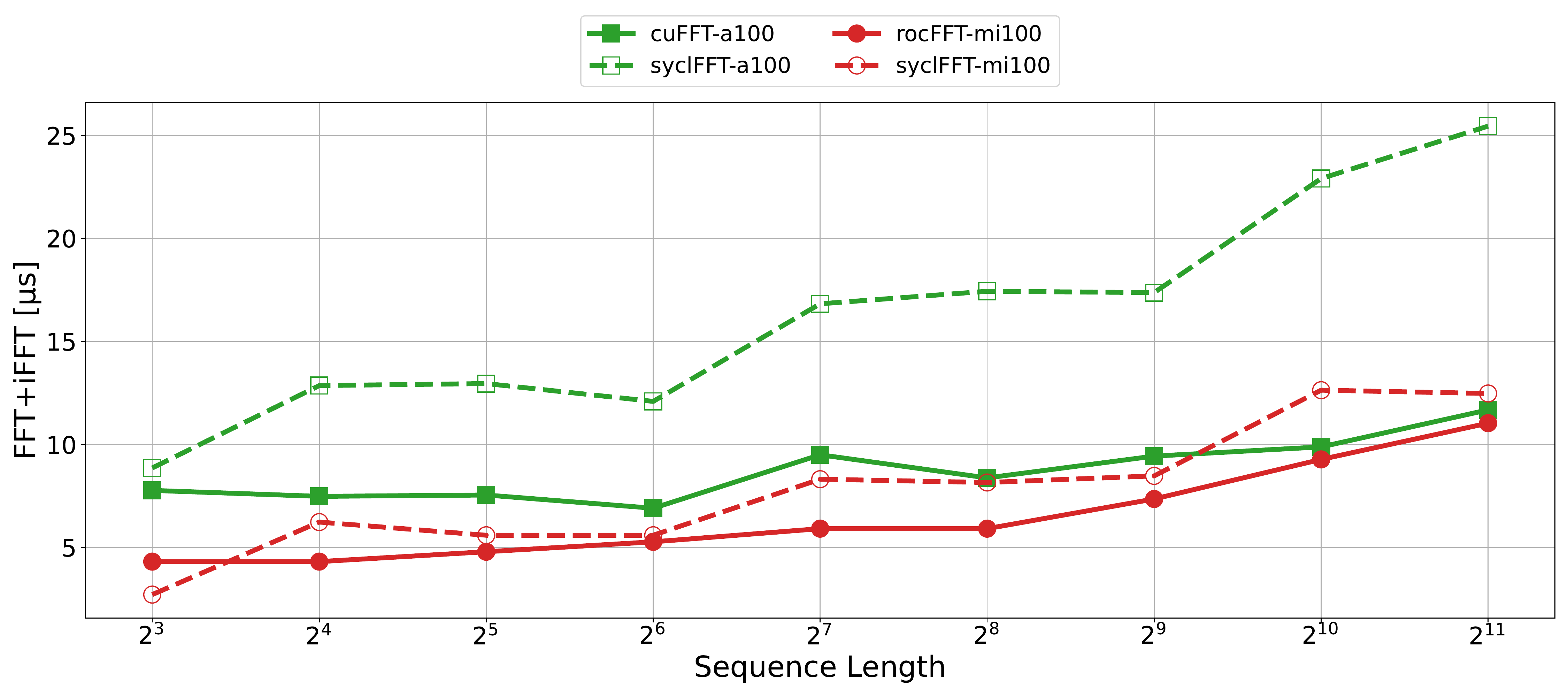}
        \caption{Optimal kernel run-times.}
        \label{fig:clock-sycl-cuhip-opt}
    \end{subfigure}
    \caption{SYCL-FFT, cuFFT and rocFFT run-times, in microseconds, on NVIDIA A100 and AMD MI-100 GPUs.
    Note the different scales of each subfigure;
    in the best case, SYCL-FFT achieves very near native rocFFT kernel performance.}
    \label{fig:clock-sycl-cuhip}
\end{figure}

Figure~\ref{fig:clock-sycl-cuhip-mean} shows the optimal total (kernel launch and execution)
time out of 1000 test runs for computing the FFT and iFFT for $f(x)=x$ on an NVIDIA A100 and
AMD MI-100 using cuFFT (solid green), rocFFT (solid red) and SYCL FFT (dashed curves).
The SYCL FFT codes for both PTX64 and HIP backends were compiled using the
\texttt{sycl-nightly/20220223} branch of Intel LLVM, built separately to target CUDA and
AMDGPU, respectively.
Compared to the analogous vendor libraries, SYCL FFT is about half as performant, as seen by comparing
the solid and dashed curves in the figure.
However, SYCL FFT total run-times are largely dominated by kernel launch overheads;
disregarding launch time and considering kernel execution time alone (dotted curves), it can be seen
that dispatch overheads are substantial, increasing the total execution time by a factor of $2$--$4$.
As such, kernel execution times do not vary significantly, with SYCL FFT performing within 30\% or better
with respect to the corresponding vendor library.
For example, Fig.~\ref{fig:clock-sycl-cuhip-opt} shows that optimal SYCL-FFT run-times (chosen
as the smallest of the 1000 test runs) do not differ largely from cuFFT or rocFFT.
Similar behavior has been observed in previous studies~\cite{goli2020towards}, where the overhead of SYCL runtime interacting with CUDA was constant ~30~$\mu$s.  

\begin{figure}[!ht]
    \begin{subfigure}[b]{0.8\textwidth}
        \centering
        \includegraphics[width=\textwidth]{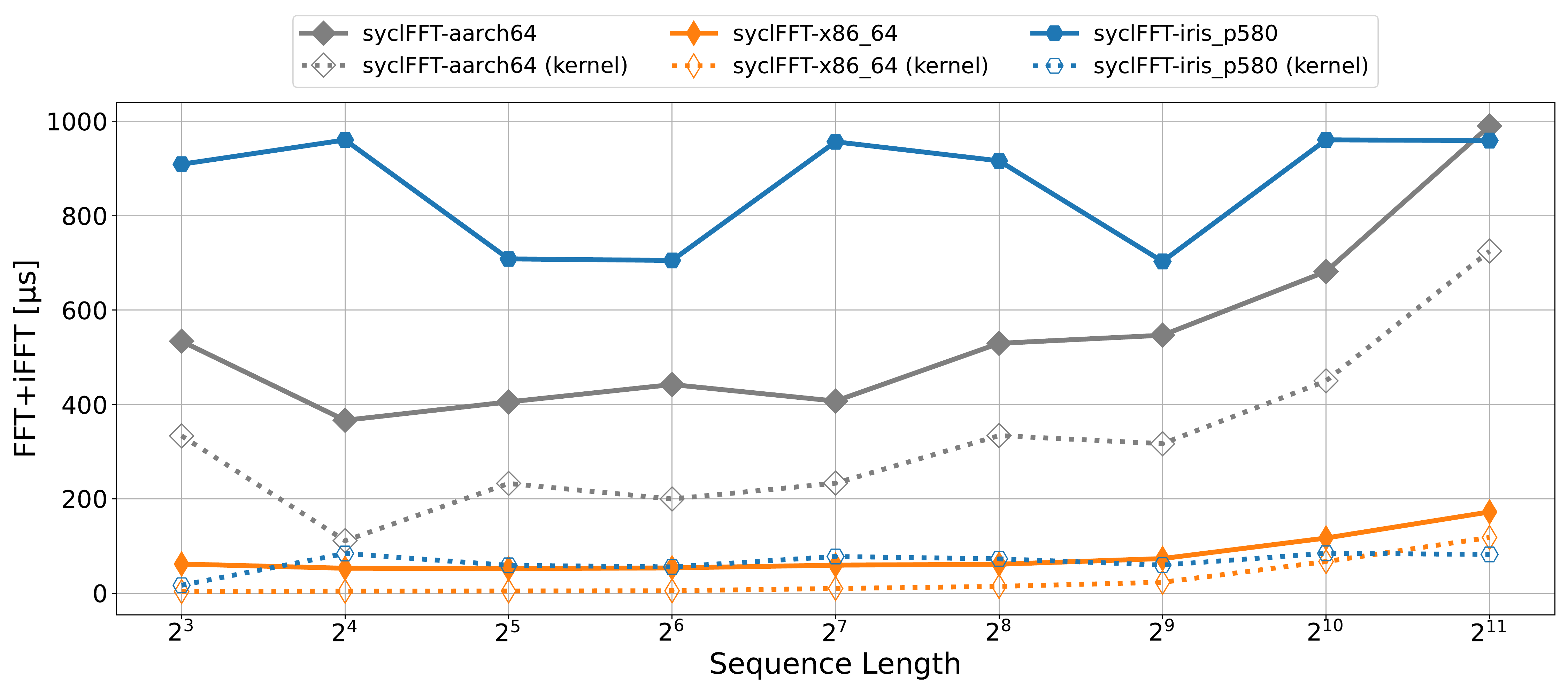}
        \caption{Mean execution time.}
        \label{fig:clock-sycl-igpu-mean}
    \end{subfigure}
    \begin{subfigure}[b]{0.8\textwidth}
        \centering
        \includegraphics[width=\textwidth]{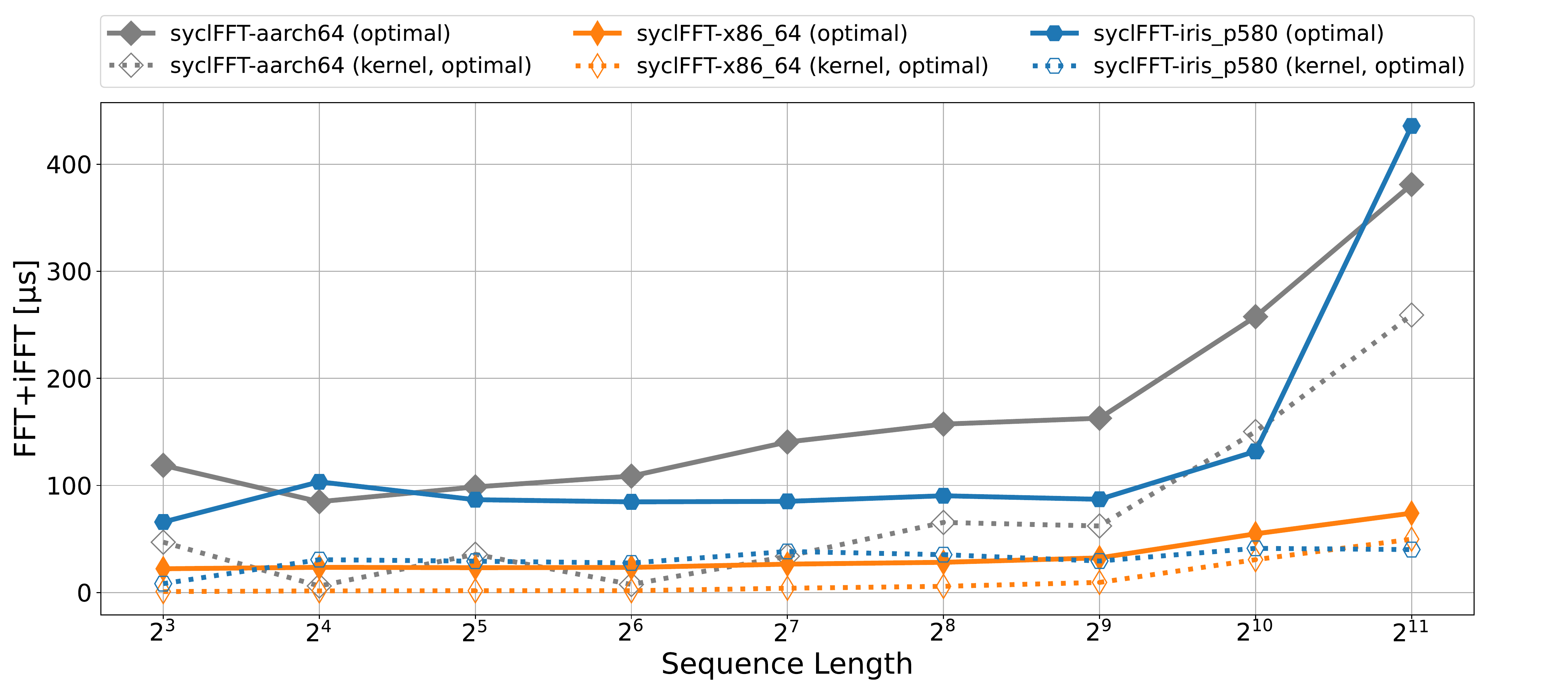}
        \caption{Optimal execution time.}
        \label{fig:clock-sycl-igpu-opt}
    \end{subfigure}
    \caption{SYCL-FFT run-times, in microseconds, on ARM, and Intel CPU and iGPU.}
    \label{fig:clock-sycl-igpu}
\end{figure}

A similar trend in launch overheads is observed for the Intel Iris P580 integrated GPU (iGPU) and ARM
Neoverse CPU, as seen in Fig.~\ref{fig:clock-sycl-igpu-mean}.
The Intel Iris P580 iGPU (blue, hexagons)---residing on the same silicon and sharing the same memory as its
host CPU---is impacted most by launch times, fluctuating by as much as 20\% between data points.
In contrast, the kernel execution times on the Intel iGPU is nearly flat across the input lengths considered.
The ARM Neoverse RISC-based CPU (gray, diamonds), which uses the POCL 1.9 prelease backend, shows the
smallest launch latency, though the kernel-only run-times are longer than would be expected.
In addition, roughly 10\% of the iterations per sequence length run on the ARM system were discarded
due to run-times exceeding the mean by an order of magnitude.
Lastly, the Intel x86\_64 CPU has the smallest overheads of all platforms considered, and displays consistent
kernel and total execution times up to an input length of $2^9$ where a linear increase occurs.
As in the cuFFT and rocFFT case, Fig.~\ref{fig:clock-sycl-igpu-opt} shows that optimal run-times
are much improved compared to the mean of 1000 test runs which have a large overall variance.

In general, we observe fair computational performance of our SYCL FFT library to cuFFT and rocFFT at
kernel level.
However, for kernels with run-times $\mathcal{O}(10)~\mu $s, the dominant contribution to total
run-times are the launching of kernels on compute devices.
These findings are summarized in Table~\ref{tab:latency}.

\begin{table}[t]
\centering
    \begin{tabularx}{0.73\columnwidth}{c|c|c}
 Device &Compiler + Backend & Launch Latency [$\mu$s] \\
 \hline
 \hline
  ARM Neoverse-N1 & ComputeCpp 2.8.0 + POCL 1.9 & 200-250 \\
 \hline
  Intel Xeon E3-1585 v5 &ComputeCpp 2.8.0 + OpenCL 3.0 &  $\sim$ 50\\
 \hline
  Intel Iris P580 &ComputeCpp + OpenCL 3.0& 650-800 \\
 \hline
  AMD MI-100 &Intel LLVM + HIP 4.2.0& $\sim$ 80 \\
 \hline
  NVIDIA A100 &Intel LLVM + CUDA 11.5.0& $\sim$ 40 (13) \\
 \hline
    \end{tabularx}
    \caption{Combinations of compiler and backend used to target the devices used in these studies, along with
    the range of corresponding kernel launch latencies.
    Shown also (in parentheses) is associated latency using \text{nvcc} with cuFFT on the NVIDIA A-100,
    obtained from an NVIDIA Nsight Compute profile.}
    \label{tab:latency}

\end{table}
\subsection{Portability and Precision}
A metric describing portability is entirely different from a metric for performance;
portability and computational throughput are arguable unrelated, and the former should pertain primarily to
reproducibility (or consistency) of the outputs.
Moreover, performance is meaningless if results are not consistent to within some margin of error.
To measure portability, we therefore consider reproducibility---that is, the level at which our portable
library agrees with platform-specific analogs.

A useful statistic for comparing distributions is the reduced $\chi^2$ test, defined as:
\begin{align}
    \chi^2_\text{reduced} = \sum_i^N \frac{(s_i - n_i)^2}{n_i} \frac{1}{\text{ndf}},
\end{align}
where the $s_i$ correspond to the SYCL FFT outputs, $n_i$ to the native library outputs in bin $i$
of their individual histograms, each having $N$ bins, and $\text{ndf}=N-1$.
Since for large $b_i$ the measurements are approximated by a Gaussian distribution, the $\chi^2$ test
statistic follows a $\chi^2$ distribution for $k=\text{ndf}$ degrees of freedom.
The probability that a set of $M$ measurements would yield a $\chi^2$ value greater than or equal to
the one obtained is referred to as the $p$-value;
a $p$-value close to unity is representative of good agreement between the $\{s_i\}$ and $\{n_i\}$.

\begin{figure}[!h]
    \centering
    \includegraphics[width=0.65\textwidth]{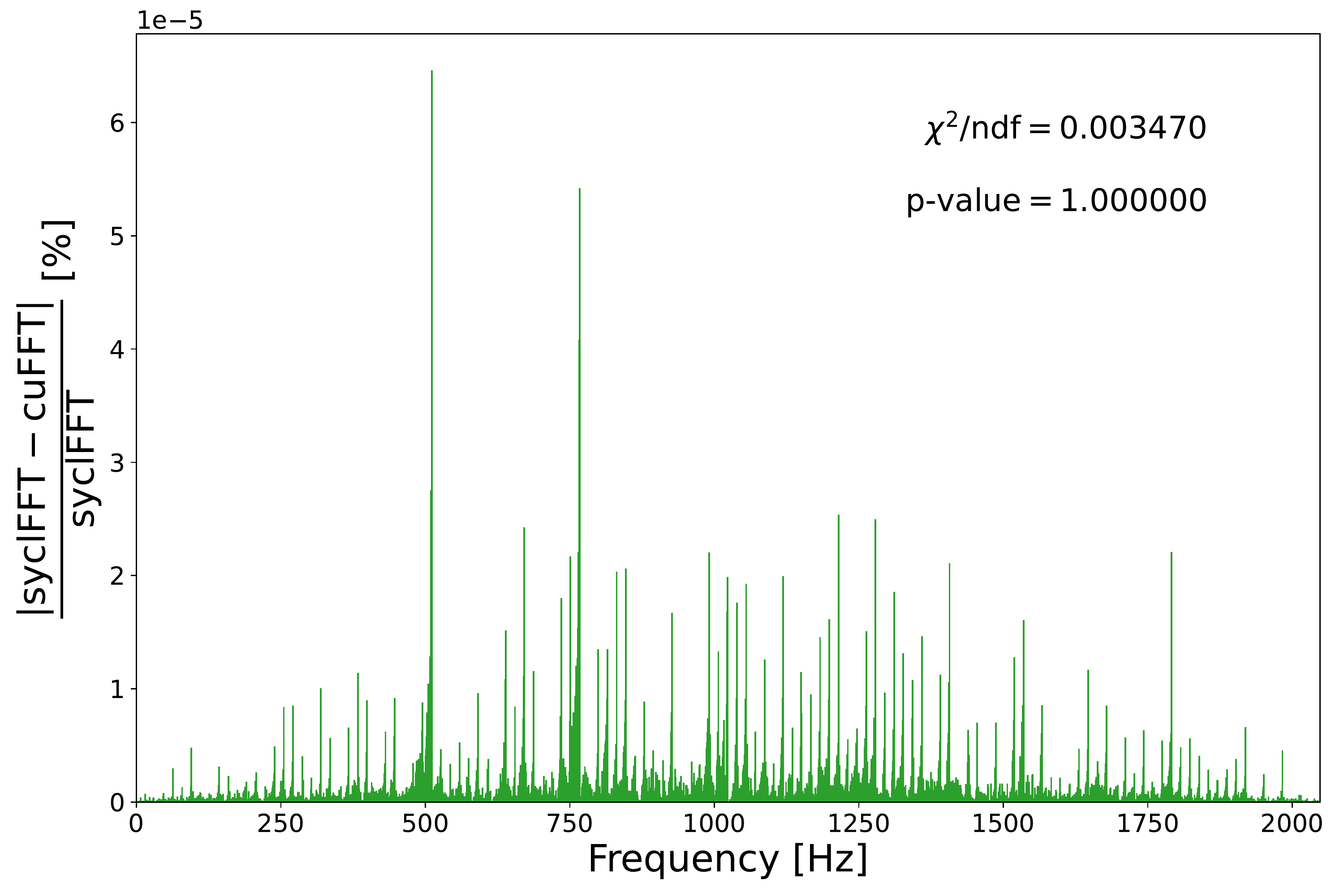}
    \caption{Absolute difference between SYCL-FFT and cuFFT outputs for a 2048 length DFT.}
    \label{fig:diff-sycl-cu}
\end{figure}

\begin{figure}[!h]
    \centering
    \includegraphics[width=0.65\textwidth]{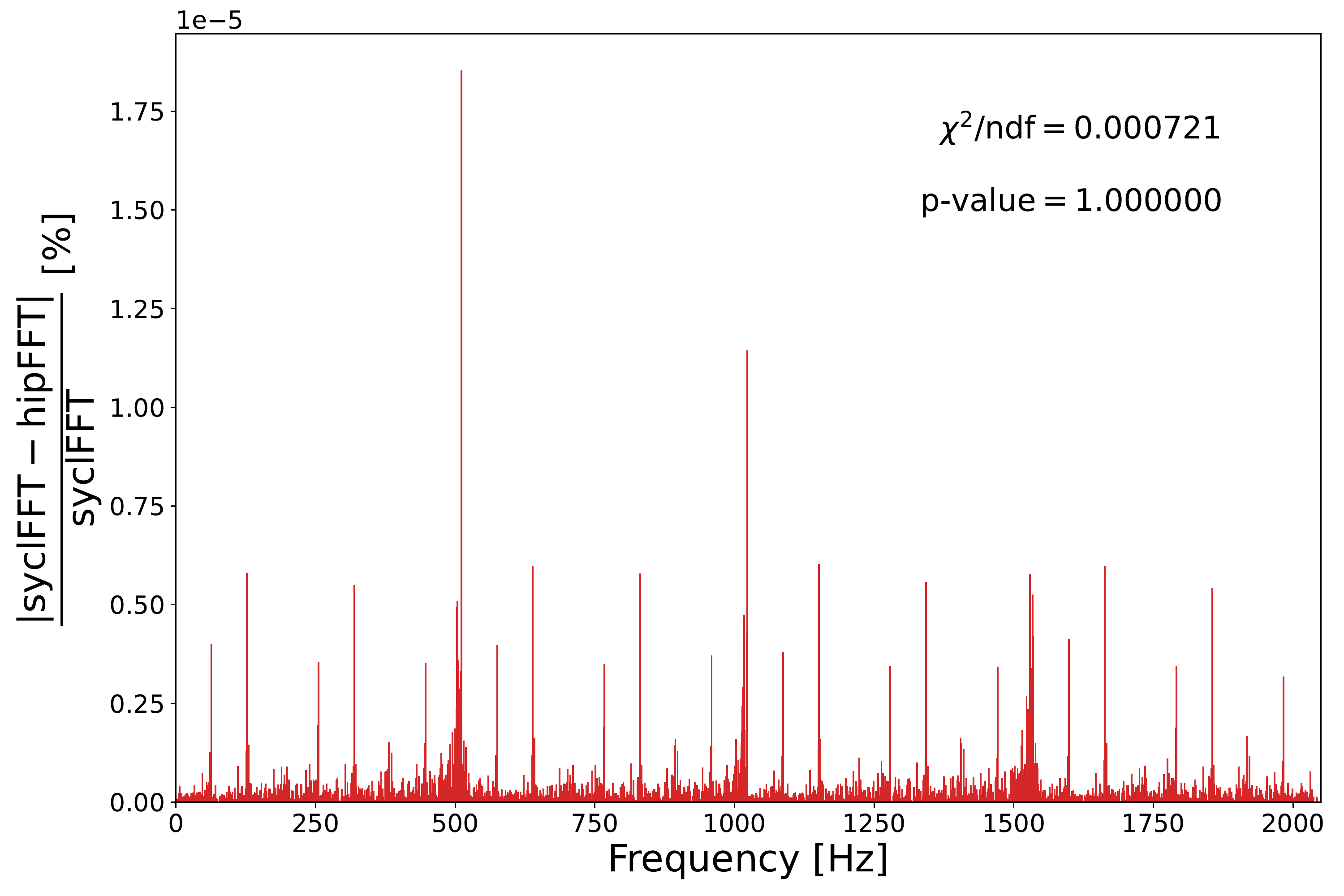}
    \caption{Absolute difference between SYCL-FFT and rocFFT outputs for a 2048 length DFT.}
    \label{fig:diff-sycl-hip}
\end{figure}

In principle, FFT algorithms are well-defined and hence different algorithms should yield exact outputs at
a given precision.
In practice, however, different rounding policies apply across devices and architectures.
Also, during application of sub-functions, textit{e.g.}, $\cos$ and $\sin$, rounding operations which are
non-associative can be applied at a low level.
Figure~\ref{fig:diff-sycl-cu} shows the ratio $|\text{syclFFT} - \text{cuFFT}|/\text{syclFFT}$, \textit{i.e.},
the difference between SYCL FFT and cuFFT output in the frequency domain for $f(x) = x$.
The calculated statistics $\chi^2/\text{ndf} = 3.47 \times 10^{-3}$ and
$p\text{-value}=1.0$ indicate a perfect agreement across the range of input sequence lengths at single precision.
The same comparison between SYCL FFT and rocFFT, shown in Fig.~\ref{fig:diff-sycl-hip}, reports a similar
level of agreement between output FFT distributions.
Since SYCL FFT is implemented to use native trigonometric functions as defined by vendors when available,
these are the expected results.
From a portable reproducibility perspective, SYCL FFT meets the desired precision.

\section{Conclusions}
In this paper we introduced SYCL-FFT, a prototype performance portable Fast Fourier Transform library
developed using the SYCL programming model.
Our library is based on Cooley-Tukey algorithms, giving $N \log n$ computational performance.
We test and benchmark SYCL-FFT on all major vendor platforms---AMD, ARM, Intel and NVIDIA---including
both CPUs and GPUs.
To evaluate our library on various hardware, we used Codeplay's ComputeCpp and Intel's open-source SYCL-enabled
LLVM compiler in conjunction with Portable Open Compute Language (POCL), Open Compute Language (OpenCL), PTX64
and HIP.
Our initial analysis sheds light on a number of important features pertaining to both the compilers and
different backends employed in this work.
Out of the box, SYCL-FFT shows a roughly 2-3x performance hit compared to vendor-optimized libraries, however,
attains the desired precision in both time and frequency domains.
Execution of over 1000 FFT exposed significant overhead costs in terms of kernel launches.
In particular, the overhead of the SYCL runtime, and especially kernel dispatch, affects the overall
performance our library.
This observation is in direct agreement with previous work~\cite{goli2020towards}; 
although the runtime overheads are significant for small problem sizes, larger problems are impacted less as the
gap between runtime and kernel execution increases.
Ongoing improvements to the various backend implementations and their offloading mechanisms can potentially close the gaps between SYCL-based---and other portable libraries---and vendor libraries. Based on these studies, AMD GPUs are most efficient for small kernels.

Future work includes expanding the library to accommodate arbitrary input sizes and 
support for multidimensional inputs.
Ultimately, we aim to provide a set of SYCL-based mathematical libraries for performance portability across all major platforms.

\begin{acks}
This work was supported by the DOE HEP Center for Computational Excellence at Brookhaven National Laboratory under B\&R KA2401045.
The authors gratefully acknowledge the computing resources provided and operated by the Joint Laboratory for System Evaluation (JLSE) at Argonne National Laboratory.
\end{acks}

\bibliographystyle{ACM-Reference-Format}
\bibliography{main}

\appendix
\section{Auxiliary Figures}
\begin{figure}[!ht]
    \begin{subfigure}[b]{0.8\textwidth}
        \centering
        \includegraphics[width=\textwidth]{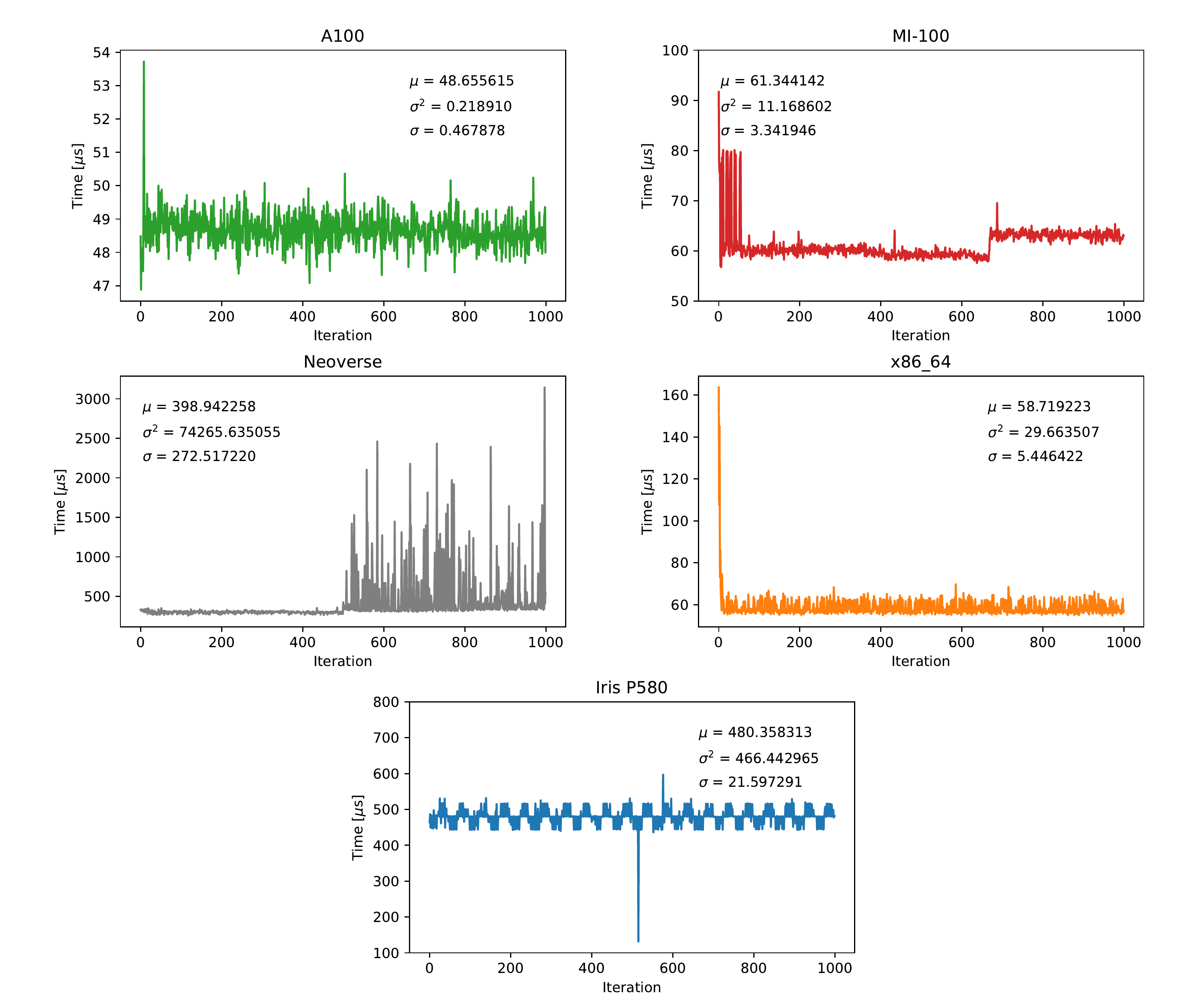}
        \caption{Input sequence length of 1024.}
        \label{fig:clock-sycl-all-1024}
    \end{subfigure} \\
    \begin{subfigure}[b]{0.8\textwidth}
        \centering
        \includegraphics[width=\textwidth]{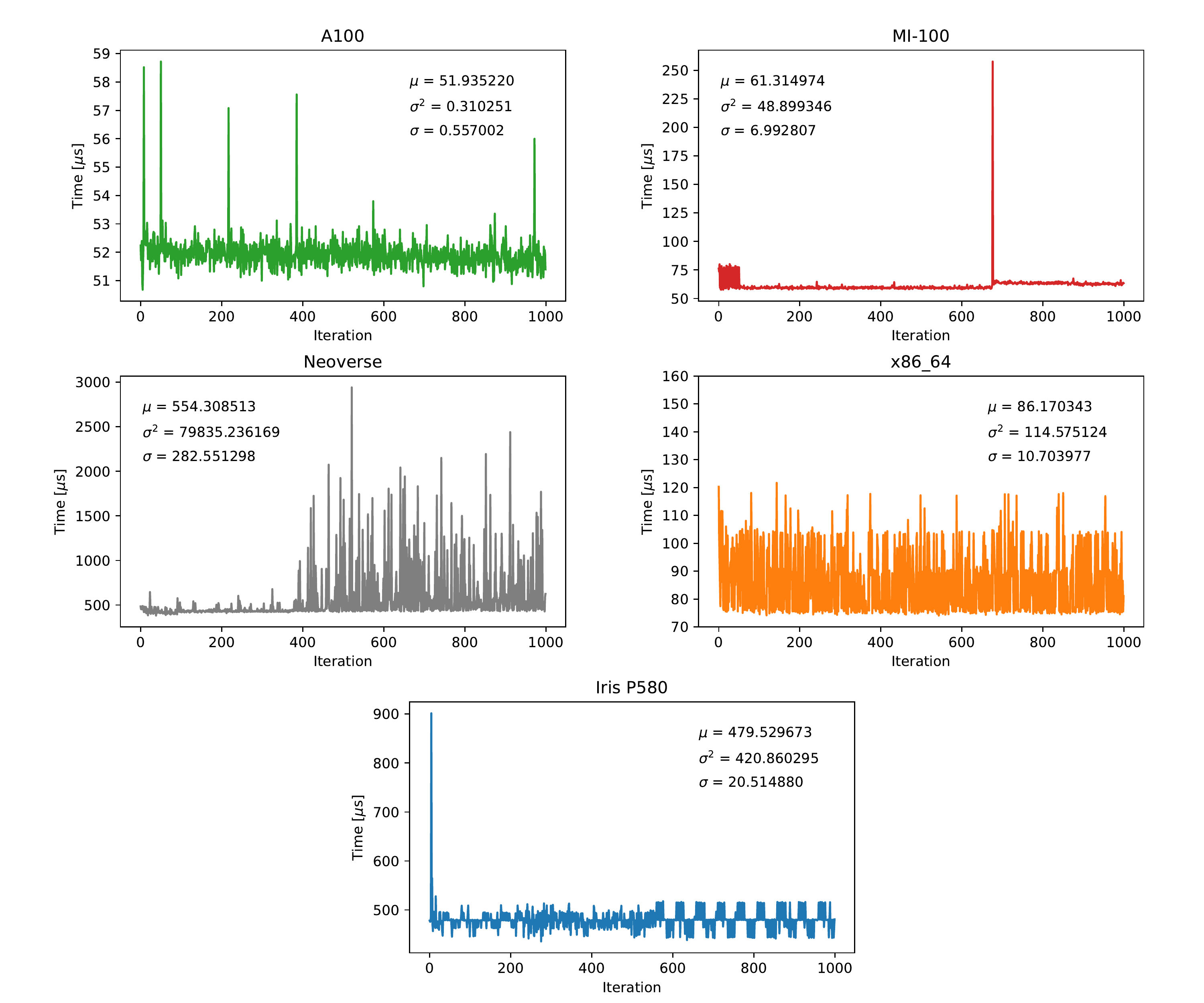}
        \caption{Input sequence length of 2048.}
        \label{fig:clock-sycl-all-2048}
    \end{subfigure}
    \caption{Distributions of 1000 combined kernel launch and execution times of SYCL-FFT across all hardware.}
    \label{fig:clock-sycl-all}
\end{figure}

Shown in Fig.~\ref{fig:clock-sycl-all} are the distributions of the combined kernel dispatch and execution times---along with their mean, variance and standard deviation---across all hardware used for benchmark studies.
A first warm-up run is discarded in all cases.
These distributions highlight the sporadic and highly fluctuating run-times measured among the various
backends.
In all cases we observe at least one outlier that negatively impacts the overall run-time of SYCL-FFT.
The A100, MI-100 and Intel CPU have mostly consistent behaviour across all 1000 tests, modulo several runs
where spikes in run-time occur.
Frequency throttling is observed in Fig.~\ref{fig:clock-sycl-all-1024} for the MI-100 after roughly 700
iterations, and around 500 iterations for the ARM Neoverse CPU.
The Intel iGPU demonstrates an interesting sinusoidal behavior, possibly due to hardware-enacted frequency
reduction and resource sharing with the host CPU.

\end{document}